\documentclass[twocolumn,twoside,prd,floatfix,letterpaper]{revtex4}

\usepackage{ifpdf}
\ifpdf 
  \usepackage[pdftex]{graphicx} 
\else 
  \usepackage{graphicx} 
\fi
\usepackage{amsmath}
\usepackage{amssymb}
\usepackage{fancyhdr}
\usepackage{color}
\usepackage{rotating}
\usepackage[colorlinks,hyperindex]{hyperref}
\usepackage{breakurl}

\def \M {{$M$}}

\begin{document} 

\title{Information, Impact, Ignorance, Illegality, Investing, and Inequality}
\author{Bruce Knuteson}
\noaffiliation

\begin{abstract}
We note a simple mechanism that may at least partially resolve several outstanding economic puzzles, including why the cyclically adjusted price to earnings ratio of the S\&P 500 index has been oddly high for the past two decades, why gains to capital have outpaced gains to wages, and the persistence of the equity premium.
\end{abstract}

\maketitle

\section{Information and Impact}
\label{sec:InformationImpact}

In United States equity markets, bid-ask spreads early in the trading day are typically larger than spreads later in the trading day~\cite{wsj2014intraday}~\footnote{The first six sentences in this article are nothing more than six restatements of one well-known fact:  in United States equity markets, spreads are wider and depths are thinner near market open than near market close.  The qualitative implications of this fact for price impact should be clear to even the dullest professional market participant.}.  The price impact of aggressive trades early in the trading day is therefore typically larger than the price impact of equally sized aggressive trades later in the day~\cite{cont2014price}.

Market makers make wider markets early in the trading day.  The market, viewed as an information aggregator, respects the information content of aggressive orders early in the day more than the information content of equally sized aggressive orders later in the day~\cite{cont2014price}.

A repeated sequence of intraday round trips -- e.g., buying in the morning and selling in the afternoon, repeated over many days -- can therefore be expected to result in net price impact in the direction of the morning trade.

\section{Ignorance}
\label{sec:Ignorance}

If some market participant (\M) performs the same round trip each day -- e.g., aggressively buying in the morning and selling in the afternoon -- \M's trading will, on average, nudge the market's midprice in the direction of his morning trading.  

If \M\ has a large, slowly varying portfolio, the mark to market gains resulting from \M's daily intraday round trip trades can exceed the cost \M\ incurs by crossing the spread twice each day.

Putting in some round numbers, suppose \M\ has \$1B in capital levered 10 times for a total equity book of \$10B.  Further suppose \M\ aggressively trades \$10M in the direction of his portfolio in the morning, crossing a typical full spread of 15~bps in many stocks, and then trades \$10M in the opposite direction at the end of the day, crossing a typical full spread of 5~bps.  \M's expected daily trading cost ($\sim$\$10K) is significantly less than the expected mark to market gain on \M's portfolio from \M's trading ($\sim$\$1M)~\footnote{\M's actual trading may be quite complicated and yet retain the essential element of our caricature:  by consistently tending to expand his portfolio in the morning and contract it near close, \M\ can create mark to market gains on a large existing book exceeding the transaction costs \M\ incurs on this trading.  Although ultimately unsustainable, under appropriate conditions this could persist for years, possibly aided by new capital allowing \M\ to further expand his book.}.  The two orders of magnitude separating these numbers leaves plenty of room for a more careful analysis to produce a qualitatively similar conclusion:  \M\ can systematically mark up his existing book by trading in the direction of his book early in the trading day and trading in the opposite direction later in the day.

The two orders of magnitude separating these numbers also leaves plenty of room for \M\ to be sloppy in his execution of this strategy.  One can imagine \M\ stumbling upon this trading style -- tending to expand his portfolio in the morning and contracting it in the afternoon -- without fully grasping its consequences; chalking his profits up to his cleverness in identifying market patterns; and either missing or choosing to ignore signs he is blowing his own bubble.  The self delusion is even easier if there are a few such \M s, sheepishly tending to hold similar portfolios, and inadvertently sharing the trading costs of marking up all of their books.

How long could this persist?  \M's investors and onlookers, seeing impressive returns and not understanding the mechanism by which they have been generated, are more likely to entrust \M\ with additional capital than to complain.  An average daily nudge of 4~bps takes seven years to push prices by a factor of two.  \M's charade can therefore plausibly last the better part of a decade \ldots\ and possibly longer, if other, fortuitous bets let \M\ wriggle out of a blowup or two.

\section{Illegality}
\label{sec:Illegality}

\M's actions are of course illegal if \M\ is aware of what he is doing, or if \M\ reasonably should be aware of what he is doing but chooses to be willfully blind.

\section{Investing}
\label{sec:Investing}

In light of the above, it is striking that the returns to the S\&P 500 index over the fifteen years spanning 1993 to 2007, inclusive, all came at the start of the trading day~\cite{cooper2008return}.

Indeed, Figure~1 of Ref.~\cite{cooper2008return} is so striking it calls for a simple explanation.  We propose such an explanation.  We propose some market participant \M, tending to trade in one direction early in the trading day and in the other direction later in the day, has had a much larger long-term effect on United States equity prices than has so far been widely appreciated~\footnote{The possibility of a systematic drift in prices should come as no great surprise.  Despite much talk about efficient markets, the stock market has no strong restoring force: a company's fundamental value, realized over decades, is difficult to determine with accuracy better than a factor of two.  Increasingly, equities trading is initiated algorithmically, with at best passing reference to fundamental value.  The daily periodicity of the stock market provides the mechanism enabling the drift, allowing a successive sequence of intraday round trip trades to leave price impact in the direction of the morning trade (Section~\ref{sec:InformationImpact}).}.

This testable proposal provides a candidate solution to several outstanding economic puzzles.  Why has the cyclically adjusted price to earnings ratio of the S\&P 500 index been oddly high for the past two decades~\cite{shiller2015irrational}?  Why have gains to capital outpaced gains to wages~\cite{piketty2014capital}?  Why has the equity premium~\cite{mehra1985equity} persisted?  The solution to all of these puzzles may lie in the systematic pattern of intraday trading of one or more specific market participants that have ratcheted equity prices to unreasonable levels over the course of the past two to three decades~\footnote{In contrast to typical discussions about the (ir)rationality of human investors, our proposal for the cause of recent United States equity bubbles is highly testable:  either one or more such \M s exist, or they do not.  Any such \M, identifiable by his footprints in trading records over many years, is likely to be a familiar name.}.

\section{Inequality}
\label{sec:Inequality}

The fact that gains to capital have outpaced gains to wages has been identified as an important contributor to recent increased wealth inequality~\cite{piketty2014capital}.  This wealth inequality could arise in significant part simply from \M's trading.

\section{Summary}
\label{sec:Summary}

We certainly hope the speculation in this article turns out to be wrong.  It would be an embarrassing shame if current prices of publicly traded companies were largely an unintended side effect of a few market participants, trading systematically in roughly the same way, day after day, for years.  It would be even worse if that trading turned out to be initiated by computer algorithms, which are very good at doing things consistently, day after day, for years.  It would be worse still if those computer algorithms were written by people with next to no knowledge of the companies being traded.  It would be an extraordinary tragedy if this has artificially contributed to undue inequality within and between nations; if the resulting aberrant price signals have encouraged millions of individuals to make imprudent decisions regarding their careers, retirement, and savings; and if the stories we have told ourselves to explain suspiciously high equity prices over the past few years have all glibly omitted the most important player(s).

For all of these reasons, the speculation in this article must surely be wrong.  It is inconceivable that regulators could have missed such an obvious pattern in United States equity markets for a full quarter century.  Even if they had, surely someone would have noticed and alerted them, and surely they would then have taken prompt and effective action.  There is no way a nation of three hundred million intelligent people could have let a few market participants produce three consecutive equity bubbles in the space of two decades.  It is ridiculous to think we could have misinterpreted the consistent actions of a few market participants as an indication of America's recovering economic strength following the financial crisis of 2008.  With so many pundits commenting continuously on the stock market, it stretches imagination to believe all of those bright, articulate people and their staffs could have missed something so basic.  The idea that ten trillion dollars of illusory equity value might have been created by the trading of a few market participants over the span of two decades is obviously, completely, utterly, and ridiculously absurd.

Then again, maybe somebody should check.

Just in case.

\bibliography{vi}

\end{document}